\documentstyle[aps,12pt]{revtex}
\input epsf
\begin{document}
\author{Shan-Ho Tsai}
\address{Institute for Theoretical Physics\\
State University of New York\\
Stony Brook, N. Y. 11794-3840, USA}
\author{and S. R. Salinas}
\address{Instituto de F\'{\i}sica\\
Universidade de S\~{a}o Paulo\\
Caixa Postal 66318\\
05315-970, S\~{a}o Paulo, SP, Brazil}
\title{Fourth-order cumulants to characterize the phase transitions of a spin-1
Ising model }
\maketitle

\begin{abstract}
Fourth-order cumulants of physical quantities have been used to characterize
the nature of a phase transition. In this paper we report some Monte Carlo
simulations to illustrate the behavior of fourth-order cumulants of
magnetization and energy across second and first-order transitions in the
phase diagram of a well known spin-1 Ising model. Simple ideas from the
theory of thermodynamic fluctuations are used to account for the behavior of
these cumulants.
\end{abstract}

\section{Introduction}

There are many attempts to characterize the order of a phase transition on
the basis of the analysis of numerical data obtained from simulations of
finite spin systems. One of the approaches to this problem consists in the
analysis of the behavior of fourth-order cumulants of physical quantities
(as the order parameter and the energy) associated with the systems under
consideration\cite{binderprl,binder}. Properties of the fourth-order
cumulants of magnetization (and energy) have been investigated in the
context of finite-size effects in magnetically (and thermally) driven
first-order transitions in Ising and Potts models\cite
{bl,clb,commentclb,etalbinder} (as well as in the case of some other systems%
\cite{bhbook,catichaetal}).

In this paper, we perform Monte Carlo simulations for the well known
Blume-Capel model\cite{beg,ts} to illustrate the behavior of the
fourth-order cumulants of magnetization and energy across first and
second-order transitions in the phase diagram of this system. We show that
it is possible to draw some conclusions from the study of relatively small
lattices. The general features of the cumulants can be accounted for by
simple arguments from the theory of thermodynamic fluctuations. In
particular, we emphasize the differences between the two types of cumulants,
and the alternative definitions of the cumulant of energy (which has not
been fully appreciated in previous investigations).

The layout of this paper is as follows. In Section 2 we define the cumulants
of a physical quantity. In Section 3 we introduce some further definitions,
and discuss some properties of the Blume-Capel model. Simulations for the
fourth-order cumulants of magnetization and energy are reported in Sections
4 and 5, respectively. We hope to have provided another example of the use
of these cumulants to characterize the order of a phase transition.

\section{Definition of the cumulants}

The cumulants of a quantity $x$ can be obtained from an expansion of the
form 
\begin{equation}
<\exp \left( x\right) >=1+<x>+\frac 12<x^2>+\frac 16<x^3>+\frac 1{24}%
<x^4>+...\;,  \label{eq1-1}
\end{equation}
where the brackets denote an average\cite{ma}. If we keep terms up to fourth
order, the logarithm of this expansion may be written as 
\begin{equation}
\ln \{<\exp \left( x\right) >\}=<x>+\frac 12Q_2+\frac 16Q_3+\frac 1{24}%
Q_4+...\ ,  \label{eq1-2}
\end{equation}
where the cumulants $Q_2$, $Q_3$, and $Q_4$, are given by 
\begin{equation}
Q_2=<x^2>-<x>^2,  \label{eq1-3}
\end{equation}
\begin{equation}
Q_3=<x^3>-3<x><x^2>+2<x>^3,  \label{eq1-4}
\end{equation}
and 
\begin{equation}
Q_4=<x^4>-3<x^2>^2-4<x><x^3>+12<x>^2<x^2>-6<x>^4.  \label{eq1-5}
\end{equation}
These cumulants can be rewritten in the more compact form, 
\begin{equation}
Q_2=<(x-<x>)^2>,  \label{eq1-6}
\end{equation}
\begin{equation}
Q_3=<(x-<x>)^3>,  \label{eq1-7}
\end{equation}
and 
\begin{equation}
Q_4=<(x-<x>)^4>-3<(x-<x>)^2>^2.  \label{eq1-8}
\end{equation}
Also, the fourth-order cumulant $Q_4$ is more often written as 
\begin{equation}
V_x(L)=1-\frac{<(x-<x>_L)^4>_L}{3{<(x-<x>_L)^2>_L^2}},  \label{eq1-9}
\end{equation}
where $L$ is the linear size of the lattice under consideration.

\section{The Blume-Capel model}

The Blume-Capel model is given by the spin Hamiltonian 
\begin{equation}  \label{eq2-1}
{\cal H}=-J\sum_{(i,j)}{S_iS_j}+D\sum_{i=1}^NS_i^2-H\sum_{i=1}^NS_i,
\end{equation}
where $S_i=+1,0,-1$, on sites $i=1,...,N$ of a Bravais lattice, and the
first sum is performed over all pairs of nearest-neighbor sites. We consider
the ferromagnetic case, with positive exchange ($J>0$) and anisotropy ($D>0$%
) parameters, which gives rise to a competition between distinct spin
orderings. In the $D/J$ versus $T/J$ space, where $T$ is the absolute
temperature, the phase diagram consists of an ordered and a disordered phase
separated by a transition line that changes character from first to
second--order at a well defined tricritical point. We use this model to
illustrate the behavior of the fourth-order cumulants.

In zero field, the ordered phase is characterized by symmetric
magnetizations, $+m_0$ and $-m_0$, with the same energy. In general, we have 
$\left\langle {\cal H}\right\rangle _L\neq 0$, and even $\left\langle {\cal H%
}^n\right\rangle _L\neq 0$, for all $n$. Therefore, the fourth-order
cumulant of energy is written as 
\begin{equation}
V_E(L)=1-\frac{<({\cal H}-<{\cal H}>_L)^4>_L}{3{<({\cal H}-<{\cal H}%
>_L)^2>_L^2}},  \label{eq2-2}
\end{equation}
where $L$ is the linear size of the lattice. As the magnetization is
symmetric, that is, $<m^n>_L=0$ for $n$ odd, the fourth-order cumulant of
the magnetization is given by 
\begin{equation}
V_m(L)=1-\frac{<m^4>_L}{3{<m^2>_L^2}}.  \label{eq2-3}
\end{equation}
It should be remarked that, as $<E^n>_L\ne 0$ for all $n$ (including odd
values of $n$), and $<m^n>_L=0$ for odd $n$, the energy and the
magnetization give rise to distinct expressions of the fourth-order cumulant.

Now we perform a preliminary Monte Carlo simulation to look at the form of
the distribution of probabilities for the magnetization (and to motivate the
choice of the distributions of energy and magnetization to be used in the
forthcoming theoretical calculations). Let us define the dimensionless
variables 
\begin{equation}  \label{eq2-4}
t\equiv \frac TJ,\qquad d\equiv \frac DJ,\text{ and}\qquad h\equiv \frac HJ,
\end{equation}
and use a Metropolis algorithm to perform Monte Carlo simulations for the
Blume-Capel model, in zero field, on a simple cubic lattice of side $L$.

\begin{figure}
\epsfxsize=5.5in
\begin{center}
\leavevmode
\epsffile{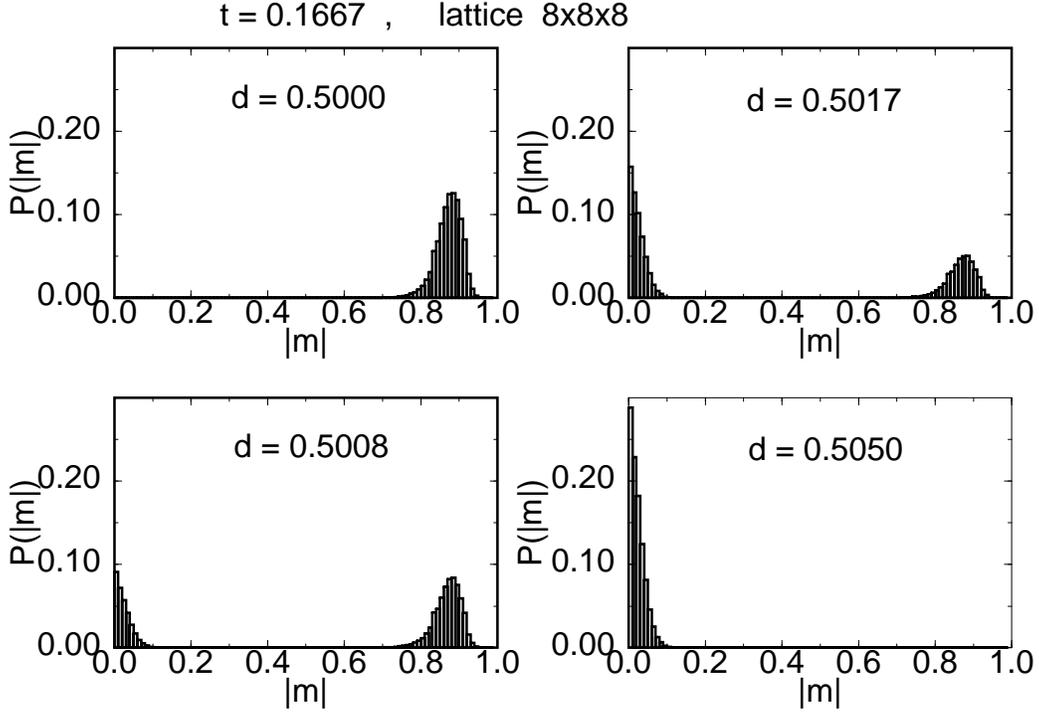}
\end{center}
\caption{Histogram of the absolute values of the magnetization, $\left|
m\right| $, across a first-order phase transition.} 
\label{fig:bc1}
\end{figure}

In Fig. (1), we show the distribution of the absolute values of the
magnetization for $t=0.1667$, $L=8$, and several values of $d$. In the top
left graph, for $d=0.5000$, $\left| m\right| $ has a unique maximum around $%
\left| m\right| =0.9$, which corresponds to an ordered phase. The bottom
left graph, for $d=0.5008$, displays two peaks (around $m=0$ and $\left|
m\right| =0.9$), which indicate the coexistence of phases ($m=0$, and $m=\pm
m_0\neq 0$). Upon increasing the value of $d$, the peak at $\left| m\right|
=0$ is enhanced, while the peak around $\left| m\right| =0.9$ is depressed
(see the top right graph, for $d=0.5017$). For even larger values of $d$,
there remains a single peak around $\left| m\right| =0$, which indicates
that the system is in the paramagnetic phase (see the graph for $d=0.5050$).
This set of graphs represents a first-order transition.

\begin{figure}
\epsfxsize=5.5in
\begin{center}
\leavevmode
\epsffile{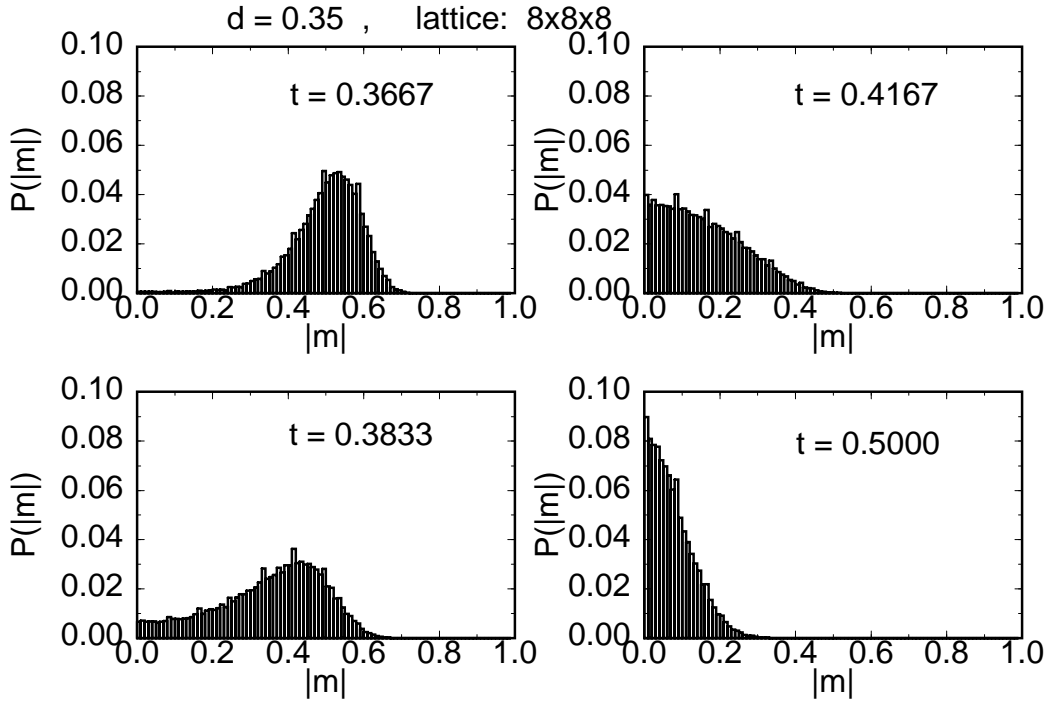}
\end{center}
\caption{Histogram of the absolute values of the magnetization, $\left|
m\right| $, across a second-order phase transition.} 
\label{fig:bc2}
\end{figure}

In Fig. (2), we show a histogram of $\left| m\right| $ across a second-order
transition, for $d=0.35$. The top left graph, for $t=0.3667$, corresponds to
the ordered phase, with a single peak of the absolute value of the
magnetization at $\left| m\right| =0.5$. As the value of $t$ is increased,
this peak moves toward $\left| m\right| =0$ and no other peak arises (see
the lower left graph, for $t=0.3833$). For higher values of $t$, $\left|
m\right| $ peaks at $\left| m\right| =0$ (see the graphs at right, for $%
t=0.4167$ and $t=0.5000$). This set of graphs illustrates a continuous phase
transition.

Figs. (1) and (2) provide the motivation for choosing a Gaussian form for
the probability of magnetization, $p(m)$. Although we are showing data for
the magnetization, similar histograms can be built for the energy, which
also give support to a Gaussian-shaped probability distribution, $p(E)$. For
large lattices $(L\to \infty )$ these distributions are expected to tend to
Dirac delta functions. The thermodynamic consistency of these assumptions
has been discussed in detail by Challa, Landau, and Binder\cite{clb} (see
also the work of Oitmaa and Fernandez\cite{of}).

Now we present separate analyses of the fourth-order cumulants associated
with magnetization and energy across first and second-order phase
transitions.

\section{Cumulants of magnetization}

In the ordered phase the distribution of probability of the magnetization
consists of two peaks around $+m_0$ and $-m_0$. For small lattices ($L$
finite), we assume the Gaussian form 
\begin{equation}  \label{eq2-5}
p(m)=\frac 12C\exp \left[ -\frac{(m+m_o)^2}{2\sigma ^2}\right] +\frac 12%
C\exp \left[ -\frac{(m-m_o)^2}{2\sigma ^2}\right] ,
\end{equation}
where $C=(2\pi \sigma ^2)^{-1/2}$ is a normalization constant, and the
parameter $\sigma $ should be inversely proportional to the lattice volume, $%
L^3$. In the $L\rightarrow \infty $ limit, we have two Dirac delta
functions, 
\begin{equation}  \label{eq2-6}
p(m)\rightarrow \frac 12\delta (m+m_o)+\frac 12\delta (m-m_o)\ .
\end{equation}

For finite lattices, with $p(m)$ given by Eq. (\ref{eq2-5}), we have $%
<m^n>=0 $, for odd values of $n$, and 
\begin{equation}  \label{eq2-7}
<m^2>=\int_{-\infty }^\infty m^2p(m)dm=\sigma ^2+m_o^2,
\end{equation}
and 
\begin{equation}  \label{eq2-8}
<m^4>=\int_{-\infty }^\infty m^4p(m)dm=3\sigma ^4+6\sigma ^2m_o^2+m_o^4.
\end{equation}
Thus, the fourth-order cumulant is given by 
\begin{equation}  \label{eq2-9}
V_m(L)=1-\frac{3\sigma ^4+6\sigma ^2m_o^2+m_o^4}{3(\sigma ^2+m_o^2)^2}.
\end{equation}
In the disordered phase, $p(m)$ has a single peak at $m_0=0$. Inserting $%
m_0=0$ in Eq. (\ref{eq2-9}), we have $V_m(L)=0$ for the disordered phase.

For infinite lattices $(\sigma \rightarrow 0)$, it is easy to see that $%
V_m(L)\rightarrow 2/3$ in the ordered phase. In the disordered phase, $%
V_m(L)\rightarrow 0$, regardless of the value of the parameter $\sigma $.
These limiting values for the infinite lattice also come from the
double-delta distribution $p(m)$, given by Eq. (\ref{eq2-6}). In this case,
in the ordered phase, we have $<m^n>=0$, for odd $n$, and $<m^n>=m_o^n$ for
even $n$. Thus 
\begin{equation}
V_m(L)\rightarrow 1-\frac{m_o^4}{3(m_o^2)^2}=\frac 23.  \label{eq2-10}
\end{equation}
In the disordered phase, $<m^n>=0$ for all $n$, hence $V_m(L)\rightarrow 0$.

\begin{figure}
\epsfxsize=3.5in
\begin{center}
\leavevmode
\epsffile{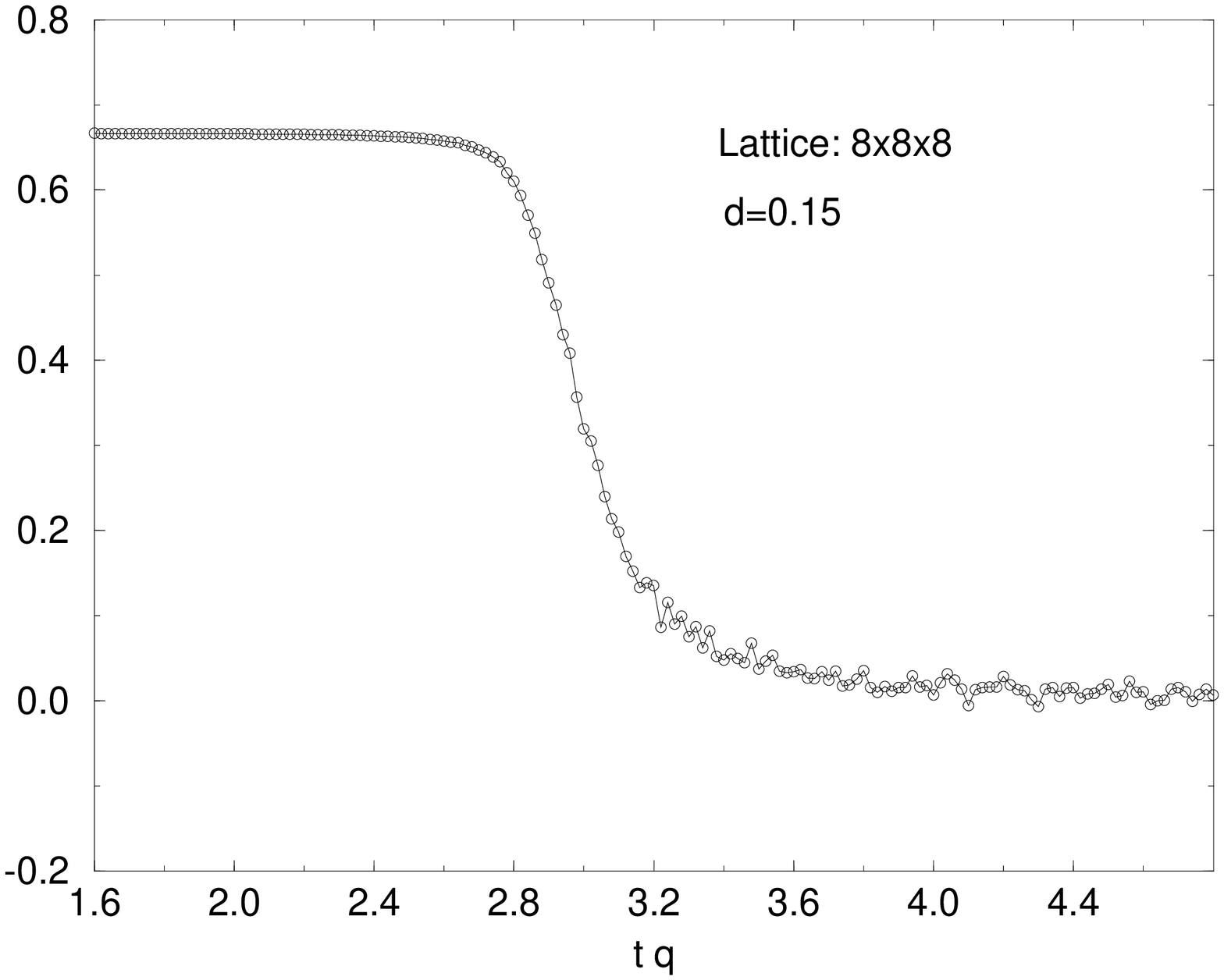}
\end{center}
\caption{Fourth-order cumulant of the magnetization versus temperature, $tq$,
across a second-order phase transition. Monte Carlo data were obtained for
the Blume-Capel model on a cubic lattice ($q=6$) of side $L=8$, for $d=0.15$. 
Averages were calculated from $50000$ Monte Carlo steps after thermalization.}  
\label{fig:vmsl8}
\end{figure}

In a second order phase transition, the two peaks of the distribution of
probabilities $p(m)$ in the ordered phase move towards each other and form a
unique peak at $m=0$ as the system passes to the disordered phase. Fig. (3)
shows the cumulant of magnetization for the Blume-Capel model in a second
order phase transition. These simulations were performed for a cubic lattice
of side $L=8$ (and coordination $q=6$), at $d=0.15$. We have run $5000$
times through the lattice to reach thermalization. Each average was then
calculated using $50000$ additional steps.

In a first-order phase transition we have the coexistence of the ordered and
disordered phases. The distribution of probabilities $p(m)$ has peaks at $%
m=\pm m_0\neq 0$, and $m=0$. For an infinite lattice, we take the
triple-delta distribution, 
\begin{equation}
p(m)=c\delta (m+m_o)+c\delta (m-m_o)+(1-2c)\delta (m)\ ,  \label{eq2-11}
\end{equation}
where $c$ is a positive constant. We then have 
\begin{equation}
<m^2>=\int_{-\infty }^\infty m^2p(m)dm=2cm_o^2,  \label{eq2-12}
\end{equation}
and 
\begin{equation}
<m^4>=\int_{-\infty }^\infty m^4p(m)dm=2cm_o^4.  \label{eq2-13}
\end{equation}
Therefore, 
\begin{equation}
V_m(L)=1-\frac{2cm_o^4}{3(2cm_o^2)^2}=1-\frac 1{6c}.  \label{eq2-14}
\end{equation}
For small values of $c$, namely $c<1/6$, the cumulant $V_m(L)$ is negative.
Similar results could have been obtained from the $L\rightarrow \infty $
limit of a Gaussian form of $p(m)$ for a finite lattice. In Fig. (4), we
show the fourth-order cumulant of the magnetization across a first-order
phase transition. The simulations were performed for a cubic lattice, with $%
L=8$, at $t=0.15$. We have run $5000$ steps through the lattice to reach
thermalization. Each average was then calculated from $50000$ additional
steps. In the ordered phase, we do have $V_m(L)=2/3$. In the disordered
phase there are very small fluctuations around $V_m(L)=0$. At the transition 
$V_m(L)$ assumes a negative value.

\begin{figure}
\epsfxsize=3.5in
\begin{center}
\leavevmode
\epsffile{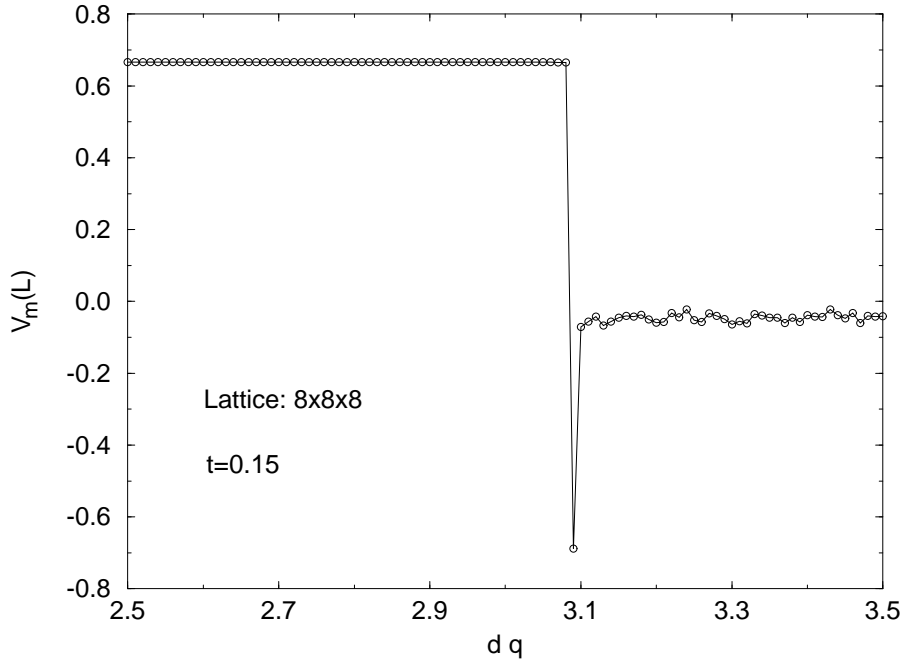}
\end{center}
\caption{Fourth-order cumulant of the magnetization versus the anisotropy, $%
dq $, across a first-order phase transition. Monte Carlo data were obtained
for the Blume-Capel model on a cubic lattice ($q=6$) of side $L=8$, at
temperature $t=0.15$. Averages were calculated from $50000$ Monte Carlo
steps after thermalization.}  
\label{fig:vmpl8}
\end{figure}

\section{Cumulants of energy}

In both ordered and disordered phases, the probability distribution of
energy has a unique peak at a certain value, which we call $E_0$. For small
lattices, we can write the Gaussian form 
\begin{equation}  \label{eq2-15}
p(E)=C\exp \left[ -\frac{(E-E_o)^2}{2\sigma ^2}\right] ,
\end{equation}
where $C=(2\pi \sigma ^2)^{-1/2}$. For large lattices ($L\rightarrow \infty $%
), we have 
\begin{equation}  \label{eq2-16}
p(E)\rightarrow \delta (E-E_o).
\end{equation}
>From Eq. (\ref{eq2-15}), we obtain 
\begin{equation}  \label{eq2-17}
<E>=\int_{-\infty }^\infty Ep(E)dE=E_o,
\end{equation}
\begin{equation}  \label{eq2-18}
<(E-<E>)^2>=\int_{-\infty }^\infty (E-E_o)^2p(E)dE=\sigma ^2,
\end{equation}
and 
\begin{equation}  \label{eq2-19}
<(E-<E>)^4>=\int_{-\infty }^\infty (E-E_o)^4p(E)dE=3\sigma ^4.
\end{equation}
Inserting these expressions into Eq. (\ref{eq2-2}), we have 
\begin{equation}  \label{eq2-20}
V_E(L)=1-\frac{3\sigma ^4}{3(\sigma ^2)^2}=0,
\end{equation}
for all values of the parameter $\sigma $ (that is, independently of the
size of the lattice used in the simulation). Using the limiting
distribution, given by Eq. (\ref{eq2-16}), we also have 
\begin{equation}  \label{eq2-21}
<E>\rightarrow \int_{-\infty }^\infty E\delta (E-E_o)dE=E_o,
\end{equation}
and 
\begin{equation}  \label{eq2-22}
<(E-<E>)^n>\rightarrow \int_{-\infty }^\infty (E-E_o)^n\delta (E-E_o)dE=0,
\end{equation}
for all $n$. Hence, $V_E(L)\rightarrow 0$, as we have already obtained.

\begin{figure}
\epsfxsize=3.5in
\begin{center}
\leavevmode
\epsffile{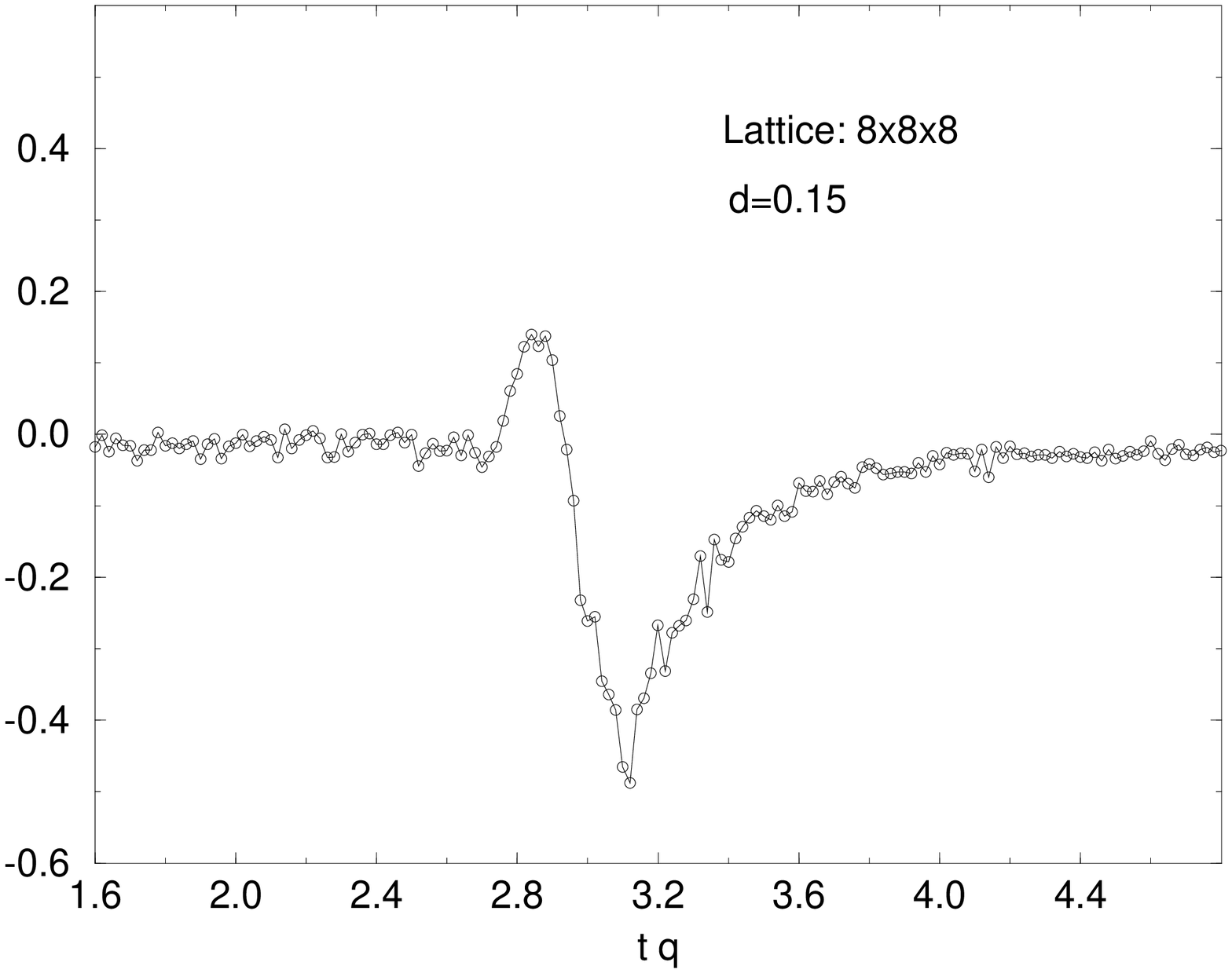}
\end{center}
\caption{Fourth-order cumulant of energy versus temperature, $tq$, across a
second-order phase transition. Monte Carlo data were obtained for the
Blume-Capel model on a cubic lattice ($q=6$) of side $L=8$, for $d=0.15$.
Averages were calculated from $50000$ Monte Carlo steps after thermalization.} 
\label{fig:vesl8}
\end{figure}

In a second-order transition the distribution $p(E)$ displays just a single
peak, that moves from an initial value $E_1$ to a final value $E_2$.
Therefore, the cumulant of energy across a second-order transition always
vanishes, independently of the lattice size. The Monte Carlo estimates of $%
V_E(L)$ for the Blume-Capel model, as shown in Fig. (5), for $d=0.15$ and
lattice size $L=8$, indicate a small maximum next to a small minimum near
the second-order phase transition. This behavior suggests that it becomes
too simple to describe the probabilities in the immediate neighborhood of a
continuous transition by a symmetric Gaussian form. This is also hinted by
the bottom left graph of Fig. (2), which is already quite asymmetric.

In a first-order phase transition there is a coexistence between the ordered
phase, associated with a distribution of energy peaked at $E_1$, and the
disordered phase, with a distribution peaked at $E_2$. Then, we can write 
\begin{equation}
p(E)=c\delta (E-E_1)+(1-c)\delta (E-E_2),  \label{eq2-23}
\end{equation}
from which we have 
\begin{equation}
<E>=\int_{-\infty }^\infty Ep(E)dE=cE_1+(1-c)E_2,  \label{eq2-24}
\end{equation}
\begin{equation}
<(E-<E>)^2>=\int_{-\infty }^\infty \left[ E-cE_1-(1-c)E_2\right]
^2p(E)dE=c(1-c)(E_1-E_2)^2,  \label{eq2-25}
\end{equation}
and 
\[
<(E-<E>)^4>=\int_{-\infty }^\infty \left[ E-cE_1-(1-c)E_2\right] ^4p(E)dE= 
\]
\begin{equation}
=c(1-c)(1-3c+3c^2)(E_1-E_2)^4.  \label{eq2-26}
\end{equation}
The fourth-order cumulant is given by 
\begin{equation}
V_E(L)=2-\frac 1{3c(1-c)},  \label{eq2-27}
\end{equation}
which becomes negative for small values of either $c$ or $1-c$. In Fig. (6),
for a cubic lattice of size $L=8$, at $t=0.15$, we show the fourth-order
cumulant associated with the energy of the Blume-Capel across a first-order
phase transition.

\begin{figure}
\epsfxsize=3.5in
\begin{center}
\leavevmode
\epsffile{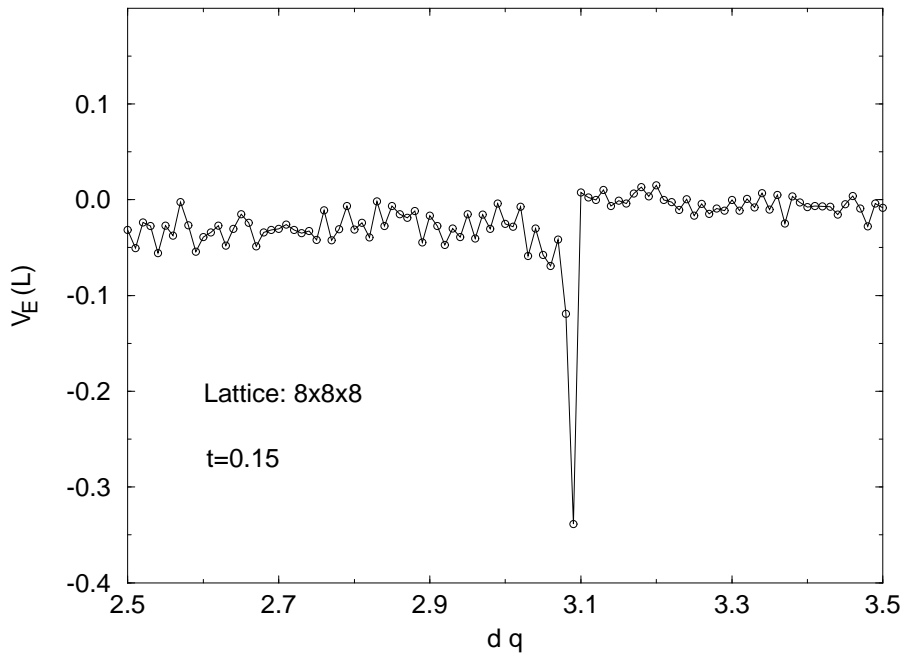}
\end{center}
\caption{Fourth-order cumulant of energy versus the parameter of anisotropy, $%
dq$, across a first-order phase transition. Monte Carlo data were obtained
for the Blume-Capel model on a cubic lattice ($q=6$) of side $L=8$, at
temperature $t=0.15$. Averages were calculated from $50000$ Monte Carlo
steps after thermalization.}  
\label{fig:vepl8}
\end{figure}

In Figs. (7) and (8), we illustrate the fourth-order cumulants of
magnetization and energy across a second-order transition. We see that the
cumulants $V_m(L)$, for different values of $L$, cross at a unique point,
which can be used to estimate the transition temperature\cite
{binder,etalbinder,bhbook,plischke}. However, it should be pointed out that
a precise location of the transition requires a detailed study of finite
size scaling, which is beyond the scope of this paper\cite
{binder,bl,clb,etalbinder,binderetal,leekost,borgs}.

\begin{figure}
\epsfxsize=3.5in
\begin{center}
\leavevmode
\epsffile{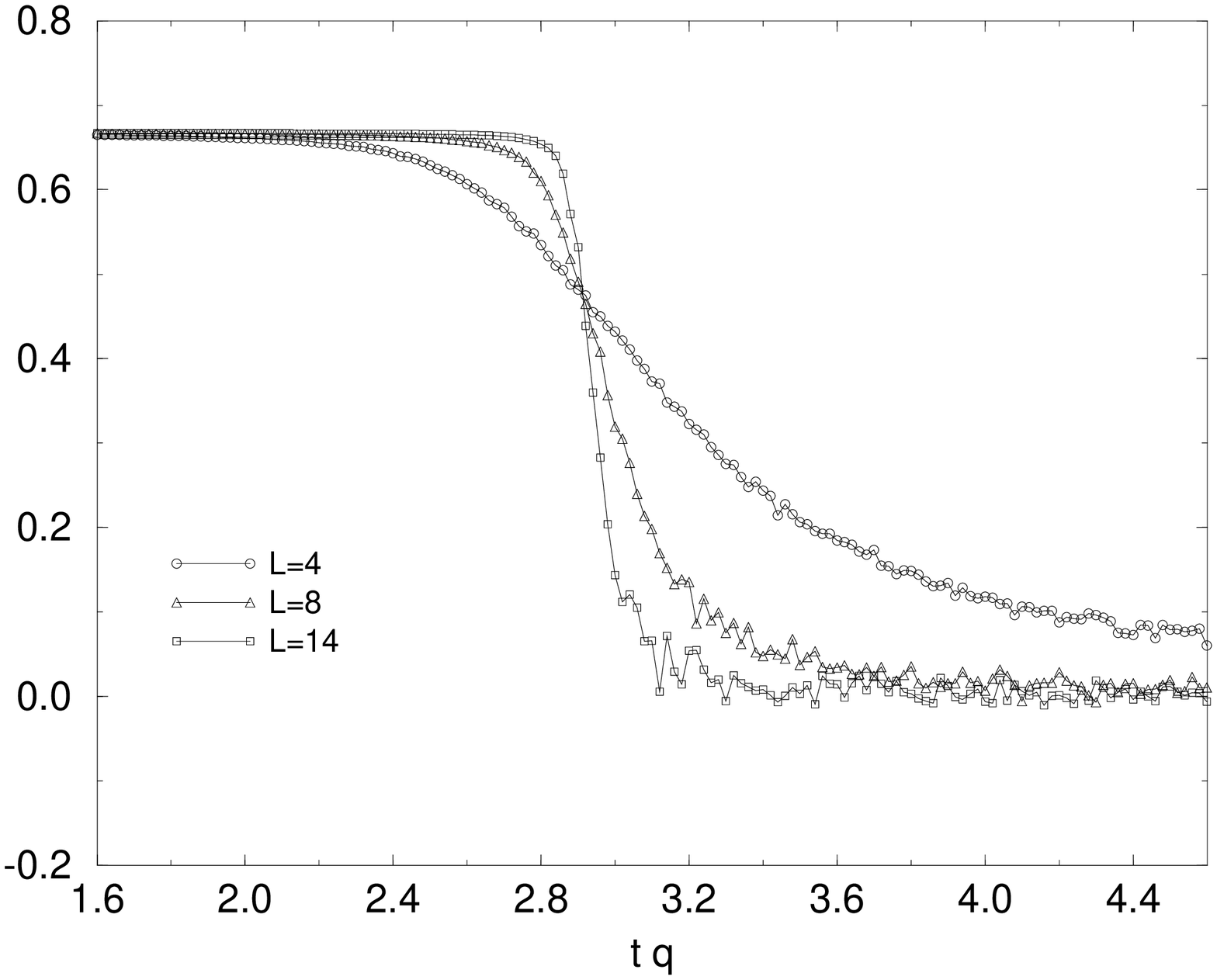}
\end{center}
\caption{Fourth-order cumulant of magnetization versus temperature, $tq$,
across a second-order phase transition ($d=0.15$), for the Blume-Capel model
on a sequence of cubic lattices ($q=6$), of lattice sizes $L=4$, $8$, and $14
$. The Monte Carlo averages were calculated from $50000$ lattice steps after
thermalization.}  
\label{fig:vmsall}
\end{figure}
\begin{figure}
\epsfxsize=3.5in
\begin{center}
\leavevmode
\epsffile{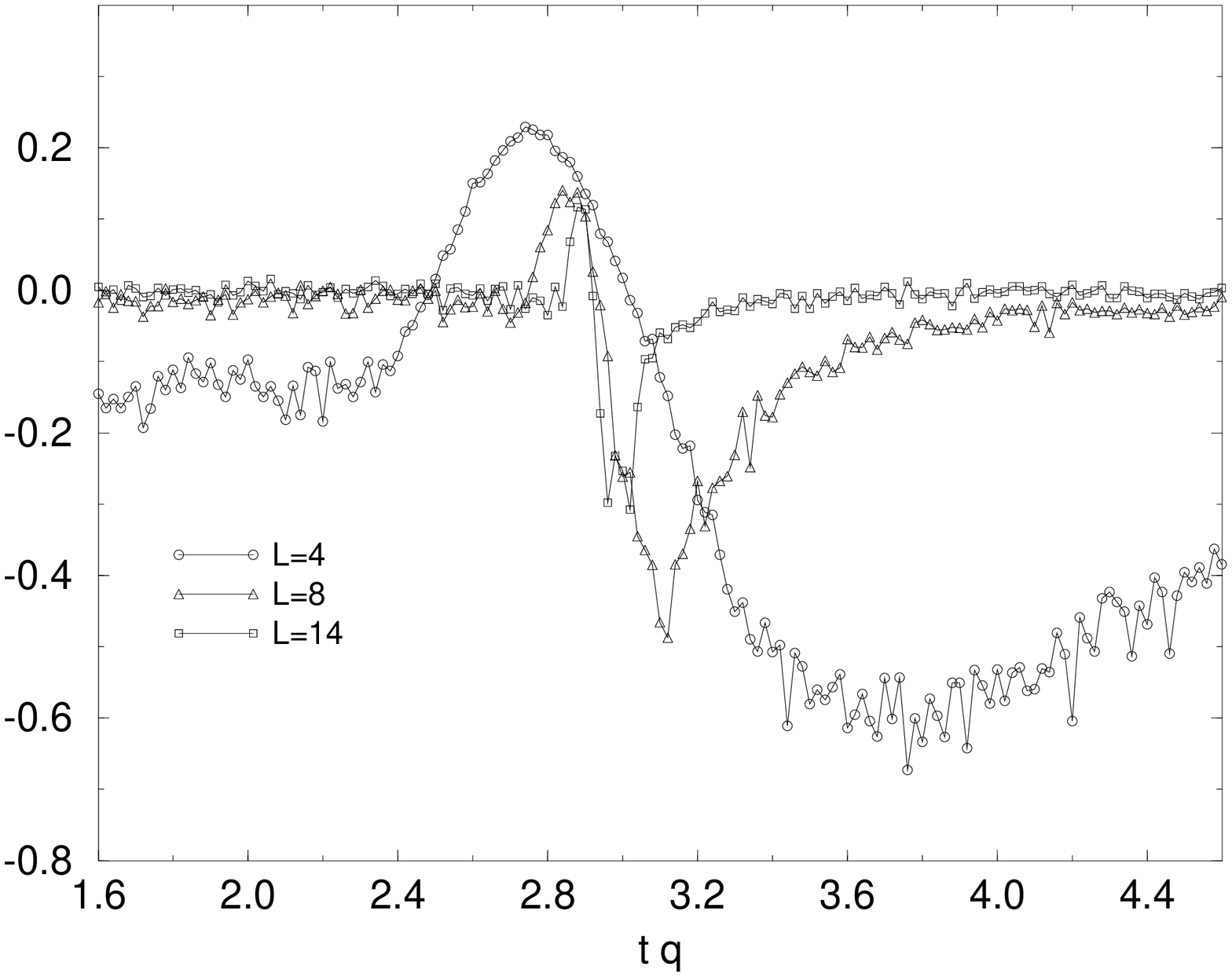}
\end{center}
\caption{Fourth-order cumulant of energy versus temperature, $tq$, across a
second-order phase transition ($d=0.15$), for the Blume-Capel model on a
sequence of cubic lattices ($q=6$), of lattice sizes $L=4$, $8$, and $14$.
The Monte Carlo averages were calculated from $50000$ lattice steps after
thermalization.}
\label{fig:vesall}
\end{figure}

Many authors use a fourth-order cumulant of the energy given by the form 
\begin{equation}
V_E(L)=1-\frac{<E^4>_L}{3<E^2>_L^2},  \label{eq2-28}
\end{equation}
instead of the connected expression of Eq. (\ref{eq2-2}). Although this may
work for Ising and Potts models\cite{jfm}, it is important to emphasize that
for the Blume-Capel model we have to use the correct definition, given by
Eq. (\ref{eq2-2}), to be able to extract the order of the phase transition
(see also the recent works of Janke\cite{janke}, and of Borgs and
collaborators\cite{borgs}).

\section{Conclusions}

We have used the spin-1 Ising model of Blume and Capel to illustrate the
feasibility of characterizing the order of a phase transition from a simple
analysis of the behavior of the fourth-order cumulants of energy and
magnetization. The general features of these cumulants can be derived from
simple arguments of the theory of thermodynamic fluctuations. In the
literature, there are two definitions of the fourth-order cumulant of
energy. We have pointed out that, in the case of systems as the Blume-Capel
model, it is important to consider the connected form of the cumulant of
energy.

{\bf Acknowledgments}

We acknowledge helpful discussions with C. E. I. Carneiro. This work was
supported by grants from Fapesp and CNPq.

\end{document}